\documentclass[prd,showpacs,preprintnumbers,floatfix,twocolumn]{revtex4}
\usepackage{amsmath}
\usepackage{graphicx}
\begin{document}
\title{Quantum Helicity Entropy of Moving Bodies}
\date{\today}
\author{Song He}
\email{she@pku.edu.cn}
 \affiliation{Institute of Theoretical
Physics, School of Physics, Peking University, Beijing, 100871,
China}
 \author{Shuxin Shao}
 \email{sxshao@mail.bnu.edu.cn}
 \affiliation{Department of Physics,
Beijing Normal University, Beijing, 100875, China}
\author{Hongbao Zhang}
\email{hbzhang@pkuaa.edu.cn}
    \affiliation{Department of Astronomy, Beijing Normal University, Beijing, 100875,
    China\\
    Department of Physics, Beijing Normal University,
    Beijing, 100875, China\\
 CCAST (World
Laboratory), P.O. Box 8730, Beijing,
   100080, China}

\begin{abstract}
Lorentz transformation of the reduced helicity density matrix for a
massive spin $\frac{1}{2}$ particle is investigated in the framework
of relativistic quantum information theory for the first time. The
corresponding helicity entropy is calculated, which shows no
invariant meaning as that of spin. The variation of the helicity
entropy with the relative speed of motion of inertial observers,
however, differs significantly from that of spin due to their
distinct transformation behaviors under the action of Lorentz group.
This novel and odd behavior unique to the helicity may be readily
detected by high energy physics experiments. The underlying physical
explanations are also discussed.
\end{abstract}

 \maketitle
Quantum information
theory is usually formulated in the framework of non-relativistic
quantum mechanics, since particles moving at relativistic speeds may
not be needed to realize the promise of quantum information process
such as quantum computation. However, relativity, especially special
relativity plays a significant role in quantum entanglement and
related quantum technology, such as teleportation. This point is
obviously justified by quantum optics, which is well established on
the basis of not only quantum theory but also special relativity in
nature\cite{Walls}. For example, most of EPR-type experiments have
been performed by photon pairs\cite{Aspect,Weihs}. In addition,
experiments of quantum teleportation have also been extensively
carried out by photons\cite{Bouwmeester,Furusawa}.

Recently, in particular, considerable efforts have been expanded on
the theoretical investigation of quantum information theory in
relativistic framework, which has gone beyond from photons to
electrons, and from explicit examples calculated in some specific
cases to general framework formulated in relativistic quantum
mechanics and even relativistic quantum field
theory\cite{Czachor1,Peres1,Peres2,Alsing1,Gingrich,Peres3,Peres4,Pachos,
Enk,Czachor2,Bergou,Alsing2,Peres5,Soo,Alsing3,Shi1,Kim,Czachor3,FM,Caban1,Kok,Ball,Lamata1,Lamata2,
Jordan1,Alsing4,Shi2,Caban2,Jordan2,Ling,Adesso}. For review, please
refer to \cite{Peres5}. A central topic in this interesting and
active research field is whether quantum entanglement is
observer-dependent. Especially, for a pure one particle state, it
has been shown that the reduced spin density matrix remains no
covariant between inertial observers with relative motion, and the
corresponding spin entropy is not an invariant scalar except in the
limiting case of sharp momenta\cite{Peres2}.

However, as inferred above, for Dirac fields, previous discussions
of relativistic quantum information theory focus primarily on
quantum entanglement between spin and momentum degrees of freedom.
On the other hand, since the helicity has an advantage in providing
a smooth transition to the massless case, it is the helicity rather
than spin that is more often under both theoretical consideration
and experimental detection in high energy physics. Although both the
helicity states and the spin states can constitute the basis of
Hilbert space of one particle, they differ in the way of unitary
transformation under the action of Lorentz group\cite{Weinberg}. As
a result, the entanglement properties for helicity differs
remarkably from those for spin after we trace out the momentum
degree of freedom. Especially, it is found that in the sharp
momentum limit, unlike the vanishing spin entropy, at small
velocities of the inertial observer the helicity entropy
demonstrates a sudden jump onto a constant value, half of the
entropy for the maximal entangled Bell states, which may be easily
detected in high energy physics experiments. Thus for both
theoretical completeness and possible implementation in high energy
physics, it is intriguing and significant to investigate quantum
entanglement between helicity and momentum in relativistic
framework. In this paper, we shall make a first step toward
investigation of this important but ignored issue.

Start with a field with positive mass $m$ and spin $\frac{1}{2}$, we
can also construct the helicity states $|p;\lambda\rangle$ as a
complete orthonormal basis for Hilbert space of one particle. The
unitary operator $U(\Lambda)$ acting on these helicity states for a
Lorentz transformation $\Lambda$ gives\cite{Weinberg}
\begin{eqnarray}
&&U(\Lambda)|p;\lambda\rangle \nonumber\\
&&=\sqrt{\frac{(\Lambda
p)^0}{p^0}}D_{\lambda'\lambda}[R^{-1}(\Lambda p)L^{-1}(\Lambda
p)\Lambda L(p)R(p)]|\Lambda p;\lambda'\rangle \nonumber\\
&&=\sqrt{\frac{(\Lambda
p)^0}{p^0}}D_{\lambda'\lambda}[B^{-1}(\Lambda p)R^{-1}(\Lambda
p)\Lambda R(p)B(p)]|\Lambda p;\lambda'\rangle \nonumber\\
&&=\sqrt{\frac{(\Lambda
p)^0}{p^0}}D_{\lambda'\lambda}[Z(\Lambda,p)]|\Lambda
p;\lambda'\rangle. \label{LT}
\end{eqnarray}
Here $R(p)$ is the rotation that carries the $z$ axis into the
direction $\mathbf{p}$, $B(p)$ is the boost from rest to the
momentum $|\mathbf{p}|$ in the $z$ direction, and $L(p)$ is the pure
boost from rest to the momentum $\mathbf{p}$. Obviously,
$L^{-1}(\Lambda p)\Lambda L(p)$ is just Wigner rotation, usually
denoted by $W(\Lambda,p)$. In addition, $D$ is the spin
$\frac{1}{2}$ irreducible unitary representation of Lorentz group.
Note that these helicity states differ in way of unitary
transformation from spin states under the action of Lorentz group,
since under Lorentz transformations spin states change according to
Wigner rotation, which is related to $Z(\Lambda,p)$ as
$Z(\Lambda,p)=R^{-1}(\Lambda p)W(\Lambda,p)R(p)$. It is of interest
by itself to investigate the implications and ramifications
resulting from this difference. However, it is another problem,
which is beyond the scope of this paper, and will be expected to be
reported elsewhere\cite{He}.

Thus a pure one particle state is represented by
\begin{equation}
|\psi\rangle=\sum_{\lambda=\pm\frac{1}{2}}\int
d^3\mathbf{p}\psi(\lambda,\mathbf{p})|p;\lambda\rangle
\end{equation}
with the normalized condition
\begin{equation}
 \sum_{\lambda=\pm\frac{1}{2}}\int
d^3\mathbf{p}|\psi(\lambda,\mathbf{p})|^2=1.
\end{equation}
It is noteworthy that this normalized state with a superposition of
various momenta represents a more physical reality since a particle
has no definite momentum in general, although for convenience the
momentum eigenstates are extensively employed in textbooks on high
energy physics and quantum field theory. Then
 the reduced helicity
density matrix associated with the above normalized state is
obtained by tracing out the momentum degree of freedom, i.e.,
\begin{eqnarray}
\rho&=&Tr_\mathbf{p}[|\psi\rangle\langle\psi|]=\int
d^3\mathbf{p}\langle\mathbf{p}|\psi\rangle\langle\psi|\mathbf{p}\rangle
\nonumber\\
&=&\sum_{\lambda,\tilde{\lambda}}\int
d^3\mathbf{p}[\psi(\lambda,\mathbf{p})\psi^*(\tilde{\lambda},\mathbf{p})|\lambda\rangle\langle\tilde{\lambda}|].
\end{eqnarray}
Here, we have used the orthonormal relation for the helicity states.
Later a Lorentz transformation $\Lambda$ changes the one particle
state to
\begin{eqnarray}
|\psi'\rangle&=&U(\Lambda)|\psi\rangle
\nonumber\\
&=&\sum_{\lambda=\pm\frac{1}{2}}\int
d^3\mathbf{p}\sqrt{\frac{(\Lambda
p)^0}{p^0}}\psi(\lambda,\mathbf{p})D_{\lambda'\lambda}[Z(\Lambda,p)]|\Lambda
p;\lambda'\rangle, \nonumber\\
\end{eqnarray}
and the reduced helicity density matrix to
\begin{eqnarray}
\rho'&=&\sum_{\lambda'\tilde{\lambda}'}\int d^3\mathbf{p}
\nonumber\\
&&\{D_{\lambda'\lambda}[Z(\Lambda,p)]\psi(\lambda,\mathbf{p})\psi^*(\tilde{\lambda},\mathbf{p})D^\dagger_{\tilde{\lambda}\tilde{\lambda}'}[Z(\Lambda,p)]|\lambda'\rangle\langle\tilde{\lambda}'|\}.\label{density}\nonumber\\
\end{eqnarray}
By Eq.(\ref{LT}), $D[Z(\Lambda,p)]$ is always an identity matrix if
$\Lambda$ is a purely spacial rotation transformation. Thus the
reduced helicity matrix is completely the same for those inertial
observers without relative motion but with different identification
of spacial direction. This property essentially stems from the fact
that the helicity $\frac{(\mathbf{p}\cdot\mathbf{J})}{|\mathbf{p}|}$
remains invariant under a purely spacial rotation
transformation\cite{spin}. Furthermore, note that any Lorentz
transformation can always be decomposed into the product of a pure
boost and a pure rotation, we next shall concentrate on what happens
to the reduced helicity matrix and the corresponding helicity
entanglement entropy when $\Lambda$ is a pure boost transformation.
Especially, taking into account that those pure boost
transformations are similarly equivalent with one another by
rotations, which means the reduced helicity density matrix only
depends on the magnitude of velocity of relative motion between
inertial observers, now we shall only need to consider pure boost
transformations along the $z$ axis.

In the special case mentioned above, set
\begin{equation}
\Lambda=\left(
          \begin{array}{cccc}
            \cosh\eta & 0 & 0 & \sinh\eta \\
            0 & 1 & 0 & 0 \\
            0 & 0 & 1 & 0 \\
            \sinh\eta & 0 & 0 & \cosh\eta \\
          \end{array}
        \right),\eta\leq 0,
\end{equation}
and
\begin{equation}
 p=m[\cosh\tau,\sinh\tau(\sin\theta\cos\phi,\sin\theta\sin\phi,\cos\theta)],\tau\geq
 0,
\end{equation}
then employing Eq.(\ref{LT}), we obtain
\begin{eqnarray}
&&D[Z(\Lambda,p)]=\left(
                    \begin{array}{cc}
                      e^{-\frac{\alpha}{2}} & 0 \\
                      0 & e^\frac{\alpha}{2} \\
                    \end{array}
                  \right)\times\nonumber\\
&&\left(
                                                      \begin{array}{cc}
                                                       \cos\frac{\beta}{2} & \sin\frac{\beta}{2} \\
                                                        -\sin\frac{\beta}{2} & \cos\frac{\beta}{2} \\
                                                      \end{array}
                                                    \right)\left(
                    \begin{array}{cc}
                      e^{i\frac{\phi}{2}} & 0 \\
                      0 & e^{-i\frac{\phi}{2}} \\
                    \end{array}
                  \right)\times
\nonumber\\
&&\left(
                 \begin{array}{cc}
                   e^\frac{\eta}{2} & 0 \\
                   0 & e^{-\frac{\eta}{2}} \\
                 \end{array}
               \right)
\left(
     \begin{array}{cc}
       e^{-i\frac{\phi}{2}} & 0 \\
       0 & e^{i\frac{\phi}{2}} \\
     \end{array}
   \right)\times\nonumber\\
   &&\left(
                                                      \begin{array}{cc}
                                                       \cos\frac{\theta}{2} & -\sin\frac{\theta}{2} \\
                                                        \sin\frac{\theta}{2} & \cos\frac{\theta}{2} \\
                                                      \end{array}
                                                    \right)\left(
                                                             \begin{array}{cc}
                                                               e^\frac{\tau}{2} & 0 \\
                                                               0 & e^{-\frac{\tau}{2}} \\
                                                             \end{array}
                                                           \right),\label{matrix}
\end{eqnarray}
where $\alpha$ and $\beta$ satisfy
\begin{equation}
\cosh\alpha=\cosh\eta\cosh\tau+\sinh\eta\sinh\tau\cos\theta,\alpha\geq
0,
\end{equation}
and
\begin{equation}
\cos\beta=\frac{\sinh\eta\cosh\tau+\cosh\eta\sinh\tau\cos\theta}{\sqrt{\sinh^2\tau\sin^2\theta+(\sinh\eta\cosh\tau+\cosh\eta\sinh\tau\cos\theta)^2}}
\end{equation}
with $\pi\geq\beta\geq 0$, respectively.

\begin{figure}[t]
\begin{center}
\includegraphics[clip=truth,width=1.2\columnwidth]{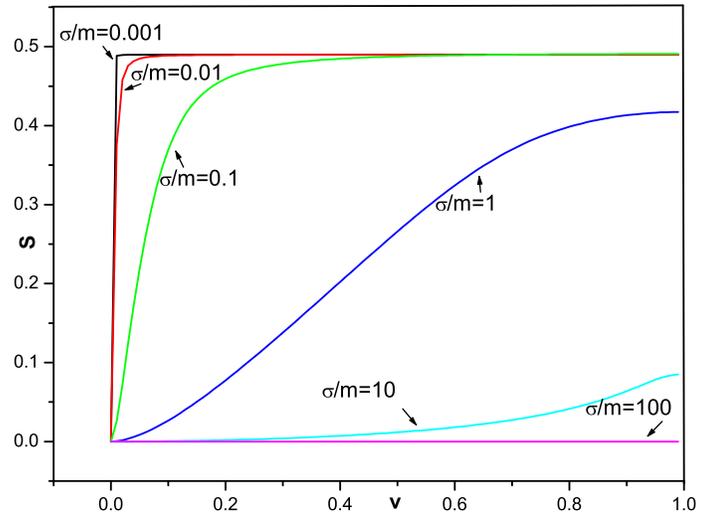}
\caption{The helicity entropy $S$ as a function of the speed $v$ of
inertial observers with respect to the original inertial reference
frame.}\label{entropy}
\end{center}
\end{figure}

As an example, consider a particle prepared in the eigenstate with
helicity $\frac{1}{2}$, i.e., right handed state, with respect to
some original inertial reference frame, which obviously implies that
the corresponding helicity entropy is zero, due to
$\psi(-\frac{1}{2},\mathbf{p})=0$. If that particle is described in
another inertial reference frame moving with velocity $v=-\tanh\eta$
along the $z$ axis of the original one, then substituting
Eq.(\ref{matrix}) into Eq.(\ref{density}), and after straightforward
but lengthy calculations, the new reduced helicity density matrix
can be obtained as
\begin{equation}
\rho'=\int
d^3\mathbf{p}\frac{|\psi(\frac{1}{2},\mathbf{p})|^2}{2}H,\label{integral}
\end{equation}
where
\begin{eqnarray}
H_{11}&=&1+\cosh\eta\sin\beta\sin\theta+\cos\beta\cos\theta,\nonumber\\
H_{12}&=&e^{-\alpha}(\sinh\eta\sin\theta+\cosh\eta\cos\beta\sin\theta-\sin\beta\cos\theta),\nonumber\\
H_{21}&=&
e^\alpha(-\sinh\eta\sin\theta+\cosh\eta\cos\beta\sin\theta-\sin\beta\cos\theta),
\nonumber\\
H_{22}&=&1-\cosh\eta\sin\beta\sin\theta-\cos\beta\cos\theta.
\end{eqnarray}
For simplicity but without loss of generality, consider in
particular the case where the wave function is a Gaussian, i.e.,
\begin{equation}
\psi(\frac{1}{2},\mathbf{p})=\pi^{-\frac{3}{4}}\sigma^{-\frac{3}{2}}e^{-\frac{\mathbf{p}^2}{2\sigma^2}},
\end{equation}
where $\sigma$ is the distribution width. However, different from
the spin case investigated in \cite{Peres2}, the later calculations,
especially the integral in Eq.(\ref{integral}), can not be carried
out analytically. Thus note that the Von Neumann entropy formula
reads
\begin{equation}
S=-tr(\rho\log_2{\rho})=-\sum\limits_{i=1,2}\rho_i\log_2{\rho_i},
\end{equation}
where $\{\rho_i\}$ are the eigenvalues of the reduced density matrix
$\rho$, we now resort to numerical methods to perform all
calculations. The corresponding results are illustrated in
Fig.\ref{entropy}.

As shown in Fig.\ref{entropy}, with the increase of speed of
inertial observers, the variation of the corresponding helicity
entanglement entropy demonstrates remarkably different behaviors
from that for the spin case investigated in \cite{Peres2}. In
particular, for the limiting case of sharp momenta which corresponds
to the small width-mass ratio $\frac{\sigma}{m}$, the helicity
entanglement entropy blows up from zero and rapidly saturates.
Speaking specifically, it arrives at a constant value(about $0.5$,
half of the entropy value for the maximal entangled Bell states) at
small velocities of observers, and then remains nearly invariant
regardless of the increase of speed of observers. While the spin
entropy always remains vanishing, as mentioned in the
beginning\cite{Peres2}. On the other hand, for the large width-mass
ratio limiting case, the resultant helicity entanglement entropy
remains zero.

Obviously this novel phenomenon differs greatly from that related to
the spin case, and unique to the helicity considered
here\cite{Peres2,Alsing1,Gingrich,Peres3,Peres4}. The key physics
underlying this seemingly odd phenomenon lies in the following fact:
With the smaller width-mass ratio $\frac{\sigma}{m}$, i.e., sharper
momentum distribution, the helicity of the prepared particle becomes
more sensitive to Lorentz boost, due to the concentration of its
momentum in a smaller neighborhood around zero. In other words, for
the smaller width-mass ratio case, the smaller speed of observers is
needed to make the flip of helicity from right to left saturate such
that the corresponding helicity entropy reaches the saturated value.

In summary, the transformation of the reduced helicity density
matrix under Lorentz group is calculated for a massive spin
$\frac{1}{2}$ particle. Especially, we have investigated the
helicity entropy for a Gaussian one particle state appealing to
numerical computation. Our results show that the helicity
entanglement entropy is not an invariant scalar, which is the same
as the previously considered spin entropy. Nevertheless, as the
speed of inertial observers increases, the specific variation of
helicity entropy demonstrates a surprisingly distinct behavior from
that of spin entropy, which essentially originates from the fact
that the helicity states differ significantly from the spin states
in transformation property under Lorentz boost. Put it another way,
unlike the spin, the helicity can be more readily flipped by
relatively small speed of observers when the momentum distribution
is sharp enough.

Associated with this unique feature of helicity, the theoretical
implications in quantum information theory and experimental
ramifications in high energy physics both need to be further
investigated. In addition, our present calculations are restricted
into a specific case in a general framework, i.e., a purely Gaussian
one particle state. A direct but non-trivial generalization is to
adopt other momentum distributions. In addition, more attractive
issues involve multi-particle state entanglement, and distillable
entanglement of mixed state for the helicity, as developed for the
spin case.

Work by SH was supported by NSFC(Nos.10235040 and 10421003). SS was
supported by LFS of BNU, and HZ was supported in part by
NSFC(No.10533010). Stimulating communications with Hengkui Wu and
Zhangqi Yin are much appreciated. Special gratitude are due to
Zhoujian Cao, Ting Chen, Huaiming Guo, Yu Lan, Haibo Yuan, Hao Wang,
and Zhi Wang for their help with numerical computations. We are also
grateful to Daniel R Terno for his helpful discussion on related
physics.


\begin{thebibliography}{99}
\bibitem{spin}Note that the reduced spin density matrix changes as $\rho'=D(\Lambda)\rho
D^\dagger(\Lambda)$ for a purely spacial rotation transformation
$\Lambda$. However, the corresponding spin entropy remains invariant
under such a rotation.
\end{thebibliography}
\end{document}